# A new production route of Gadolinium-148 for use in Radioisotope Thermoelectric Generators

[1]Ozan Artun
[1]Zonguldak Bülent Ecevit University, Zonguldak, Turkey
e-mail: *ozanartun@beun.edu.tr*; *ozanartun@yahoo.com*

**Abstract:** The production of Gd-148 had challenges for available methods in the literature, such as its high cost and low production amount. Therefore, we recommended a new production route of Gd-148 on natural Sm and Eu targets via particle accelerators. For this aim, we calculated and simulated cross-section, activity, the yield of product, and integral yield curves for 21 different nuclear reaction processes under certain conditions. Based on the obtained results, we proposed the radioisotope Gd-148 to use Radioisotope Thermoelectric Generators for deep space and planetary explorations in spacecraft and space-probes as a suitable energy source, instead of Pu-238.

## 1. Introduction

Nuclear batteries that depend on ionizing radiation to provide heat and ion production have been developed for applications of deep space researches, planetary explorations, and microelectronic technology *e.g.* the radioisotope thermoelectric generator (RTG), which converts heat power into electricity via the Seebeck effect. In a few decades, the National Aeronautics and Space Administration (NASA) has taken important steps to develop RTG which is the most advanced Multi-Mission Radioisotope Thermoelectric Generator (MMRTG) developed by NASA. MMRTG has been basically used by NASA's deep space missions to ensure electrical power to space-crafts and space-probes as shown in Fig. 1a, where Perseverance Rover in Mars mission has an MMRTG. According to the type of chosen mission, the selection of radioisotope providing heat propounds vital importance for concerning application due to the half-life and the specific power density of radioisotope. For example, as seen in Fig. 1, Pu-238 has been used by NASA for General Purpose Heat Source (GPHS) in MMRTGs due to long half-life ($T_{1/2}$=87.7 y), low radiation level (decay type: major in alpha), and high power density (0.55 watt/gram) [1-10]. In contrary to the suitable properties, the production of Pu-238 has important challenges because of Pu-238's obtaining cost and little amount in the world. However, instead of Pu-238, a new radioisotope with similar properties can be recommended, such as Gd-148, which is half-life $T_{1/2}$=74.6 y, Power=0.61 watt/gram, Decay=100% alpha [10].

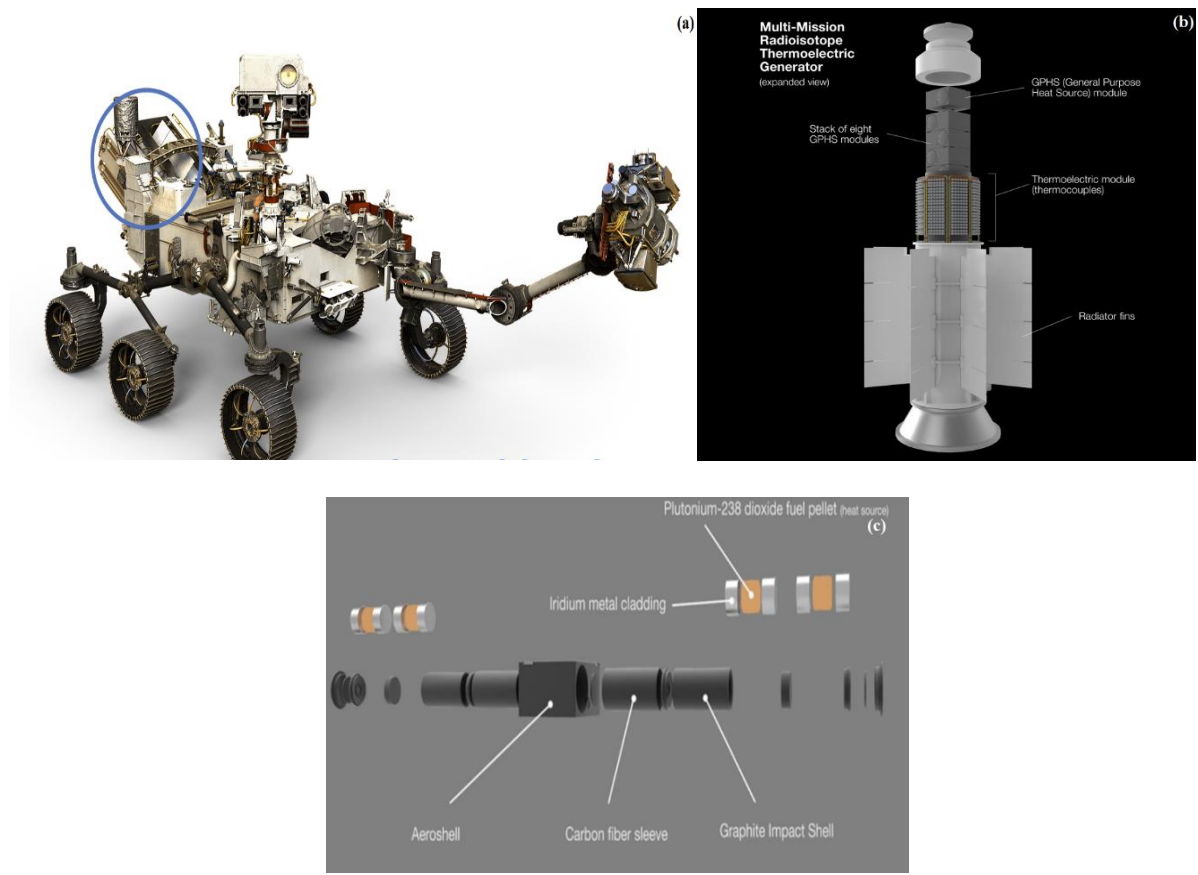

**Fig. 1. (a)** MMRTG (blue circle) of Perseverance Rover in Mars mission **(b)** A view of MMRTG **(c)** A view of GPHS [6-8].

The objective of this work is to ensure a cheaper and more suitable method for the production of Gd-148 than other production ways in the literature. Therefore, the production of Gd-148 is estimated by particle accelerators on Sm-144, Sm-147, Sm-148, Sm-149, Sm-150, Sm-152, Sm-154, and Eu-151, Eu-153 targets [11,14]. It is obvious that the radioisotope Gd-148 has ideal properties for RTGs because it has a long half-life (74.6 years), low radiation level (alpha decay mode), and high power density (0.55 watt/gram) [10]. There are a few methods in the literature for the producion of Gd-148 *e.g.* using Ta, W, and Au targets in proton-induced reactions. But, for these targets, the cross-sections of reactions are low, and proton incident energies reach up to 28 GeV [15-18]. In addition to the production methods mentioned above, different experiments or calculations for the production of Gd-148 are available in the literature. However, all of them are insufficient due to low cross section values and enormous incident particle energies (up to 208 GeV) in the reaction processes [19]. These circumstances lead to high costs and challenges in the production of Gd-148. Instead, Gd-148 can be directly produced by particle accelerators on the natural Sm and Eu targets in the energy range 1-100 MeV for proton, deuteron, triton, helium-3, and alpha-induced reactions.

For this aim, in this work, we have assessed the production of Gd-148, which has potential for use in RTGs, on Sm, Eu targets via particle accelerators. The activities and the yields of product for the productions of these isotopes are simulated under some conditions *e.g.* irradiation time of 24 h, the incident beam current of 1 mA, and the cooling time of 24 h. Moreover, the cross-sections of each reaction process are calculated in the energy region between 1 MeV and 100 MeV. Based on the obtained results, the integral yield curves of appropriate reaction processes for the production of Gd-148 to be able to use as an energy source in RTGs are calculated as dependent on incident particle energy (projectile energy).

## 2. Material and Methodology

Based on Sm and Eu targets, Gd-148, which is the suitable half-life and specific power densities to use in RTGs, might be produced by nuclear reaction processes via a particle accelerator. However, to produce this nucleus, the simulations of the reaction processes were carried out under particular circumstances to be determined the activities and the yields of product in energy region $E_{particle}=1\rightarrow100$ MeV for each reaction. The Sm and Gd targets were supposed to the uniform thicknesses and density during the irradiation time of all reaction processes and above 99% purity of targets. The reaction processes include no loss in the activity or yield. That's why the simulation of the reaction processes does not contain any superiority. Based on incident particle energy and particle beam current, it has been noted that the target isotopes in each reaction process involve appropriate target thicknesses. The effective targets of Sm and Eu isotopes and the most important results of reaction processes are clearly presented in Table 1 in which the effective targets alter between 0.172 cm and 1.696 cm. Moreover, in irradiation Sm and Eu targets, the simulation of each nuclear reaction process has been carried out by particle accelerator with the particle beam current of 1 mA in the energy range $E_{particle}$=100→1 MeV and irradiation time of 24 h. Besides, the cooling time is 24 h. Under circumstances mentioned above, the yields of product and activities of Gd-148 on Sm and Eu targets were simulated for 21 different reaction processes. To assess the production of Gd-148 on Sm and Eu targets in the nuclear reaction processes, we have calculated the cross-section of reaction processes via Talys 1.9 code, and the cross-section curves of reactions are shown in Fig. 2 as a function of projectile energy. In the cross-section calculations of nuclear reactions, we used two-component exciton model including pre-equilibrium reaction (PEQ) mechanism, and this model can be expressed in terms of PEQ population (P), the emission rate ($W_k$) as follows [20]:

$$\frac{d\sigma_k^{PE}}{dE_k} = \sigma^{CF} \sum_{p_\pi=p_\pi^0}^{p_\pi^{max}} \sum_{p_\nu=p_\nu^0}^{p_\nu^{max}} W_k(p_\pi, h_\pi, p_\nu, h_\nu, E_k) \tau(p_\pi, h_\pi, p_\nu, h_\nu) \times P(p_\pi, h_\pi, p_\nu, h_\nu), \quad (1)$$

where $E_k$ and $\sigma^{CF}$ represent the emission energy and the composite-nucleus formation cross section. Furthermore, particle (p) and hole (h) number are given p=$p_\pi$+$p_\nu$ and h=$h_\pi$+$h_\nu$ by proton (π) and neutron (ν) particles.

Regarding the level density model in two-component exciton model, we exploited the Fermi gas model with constant temperature (CTM), and total level density can be given by the following equation:

$$T\rho_F^{tot}(E_M)exp\left[\frac{-E_M}{T}\right]\left(exp\left[\frac{E_U}{T}\right] - exp\left[\frac{E_L}{T}\right]\right) + N_L - N_U = 0 \tag{2}$$

where U and $E_M$ are the effective excitation energy and an iterative term, $N_L$ and $N_U$ are constant parameters obtained from tables and databases. In addition to the cross-section calculations, the simulations of the activities and the yields of product for all reaction processes have been performed by Talys code under certain conditions. The activity ($A_j$) of the produced nucleus $j$ as a function of irradiation time ($t$) in terms of the decay rate of $j$, $\lambda_j = ln2/T_j^{1/2}$.

$$A_j(t) = \lambda_j N_j(t), \tag{3}$$

where $N_j(t)$ is represents by

$$N_j(t) = N_T(0)R_{T \to j}t \tag{4}$$

Thence, the activity is expressed as follows:

$$A_j(t) = \lambda_j N_T(0)R_{T \to j}t, \tag{5}$$

where the production rate $R_{T \to j}$ can be expressed in terms of the beam current ($I_{beam}$), the residual production cross-section ($\sigma_j^{rp}$), the projectile charge number ($z_p$) and the active target volume ($V_{target}$):

$$R_{T \to j} = \frac{I_{beam}}{z_p q_e} \frac{1}{V_{tar}} \int_{E_{back}}^{E_{beam}} \left(\frac{dE}{dx}\right)^{-1} \sigma_j^{rp}(E)dE, \tag{6}$$

Additionally, the integral yields of nuclear reaction processes have been calculated by the following equation:

$$A = \frac{N_A H}{M} I(1 - e^{-\lambda t}) \int_{E_1}^{E_2} \frac{\sigma(E)dE}{\left(\frac{dE}{d(\rho x)}\right)} \tag{7}$$

where $\left(\frac{dE}{d(\rho x)}\right)$ is the mass-stopping power of target material, $N_A$, $\sigma(E)$ and M represent the Avagadro number, the reaction cross-section and the mass number of the target material.

Furthermore, to calculate the integral yield, the mass-stopping powers of target materials need in Eq. (7). Therefore, we utilized X-PMSP program [11, 21] for the calculation of the mass-stopping power. The formulation of the mass stopping power can be expressed in terms of the atomic number of the incident particle (z), the density effect correction (δ), the velocity of the incident particle β(v/c), the mean ionization potential of the target material (I):

$$\left(\frac{dE}{\rho dx}\right) = 0.3071 \frac{Zz^2}{A\beta^2}\left[13.8373 + ln\left(\frac{\beta^2}{1-\beta^2}\right) - \beta^2 - ln(I) - \frac{\delta}{2}\right], \qquad (8)$$

where A and Z represent the mass number and proton number for the target material. Each of the variables can be found out in Artun's previous works in detail [11-14].

**Table 1:** The obtained data for the production of Gd-148 radioisotope.

| Reactions | Effective target thicknesses (cm) | Maximum cross-section (mb) | Energy value at maximum cross-section (MeV) | Activity at 24 h irradiation of target (MBq) | Maximum integral yield (GBq/mAh) | Suitable energy range (MeV) |
|---|---|---|---|---|---|---|
| $^{144}Sm(\alpha,\gamma)$ | 0.172 | 0.52 | 18 | $3.44 \times 10^{-4}$ | $13.8 \times 10^{-3}$ | 17→35 |
| $^{147}Sm(^3He,2n)$ | 0.214 | 12.60 | 22 | 0.98 | 1.68 | 22→45 |
| $^{147}Sm(\alpha,3n)$ | 0.175 | 1265.61 | 37 | 56.55 | 96.20 | 30→50 |
| $^{148}Sm(^3He,3n)$ | 0.216 | 29.56 | 28 | 3.11 | 5.19 | 23→60 |
| $^{148}Sm(\alpha,4n)$ | 0.176 | 1288.39 | 46 | 76.03 | 126.24 | 40→60 |
| $^{149}Sm(^3He,4n)$ | 0.217 | 29.49 | 38 | 4.38 | 7.29 | 33→65 |
| $^{149}Sm(\alpha,5n)$ | 0.178 | 1176.47 | 54 | 82.44 | 136.41 | 45→70 |
| $^{150}Sm(^3He,5n)$ | 0.219 | 28.58 | 47 | 5.22 | 8.67 | 40→80 |
| $^{150}Sm(\alpha,6n)$ | 0.179 | 1125.80 | 62 | 86.80 | 143.25 | 55→75 |
| $^{152}Sm(^3He,7n)$ | 0.221 | 19.94 | 60 | 4.13 | 6.83 | 55→90 |
| $^{152}Sm(\alpha,8n)$ | 0.181 | 944.09 | 77 | 75.03 | 122.83 | 70→90 |
| $^{154}Sm(^3He,9n)$ | 0.224 | 15.22 | 69 | 2.51 | 4.12 | 65→95 |
| $^{154}Sm(\alpha,10)$ | 0.184 | 912.01 | 89 | 67.11 | 109.34 | 83→100 |
| $^{151}Eu(d,5n)$ | 1.674 | 339.68 | 59 | 1036.89 | 1756.31 | 30→80 |
| $^{151}Eu(t,6n)$ | 1.248 | 73.61 | 54 | 111.87 | 186.92 | 36→70 |
| $^{151}Eu(^3He,x)$ | 0.312 | 449.60 | 57 | 133.05 | 217.22 | 42→90 |
| $^{151}Eu(\alpha,x)$ | 0.255 | 175.51 | 75 | 92.99 | 38.35 | 70→90 |
| $^{153}Eu(d,7n)$ | 1.696 | 213.15 | 85 | 670.22 | 1141.49 | 45→95 |
| $^{153}Eu(t,8n)$ | 1.264 | 38.26 | 66 | 62.11 | 103.28 | 55→85 |
| $^{153}Eu(^3He,x)$ | 0.316 | 275.86 | 72 | 79.51 | 129.26 | 55→100 |
| $^{153}Eu(\alpha,x)$ | 0.259 | 139.72 | 90 | 72.41 | 25.15 | 85→100 |

## 3. Results and Discussions

To produce Gd-148, the cross-section curves of 21 different reaction processes with charged particle-induced reactions on natural Sm and Eu targets are presented in Figure 2 as a function of projectile energy. It is clearly visible that the reactions on targets with natural Sm isotopes include He-3 and alpha-induced reactions, and the maximum cross-section curves are alpha-induced reactions. Unfortunately, the cross-section values of He-3 induced reactions are too low as compared alpha-induced reactions on Sm targets. The cross-section values of $^{148}$Sm(α,4n) and $^{147}$Sm(α,3n) reaction processes, which are the highest cross-section curves, for the production of Gd-148 are ~1288 mb and ~1265 mb, respectively.

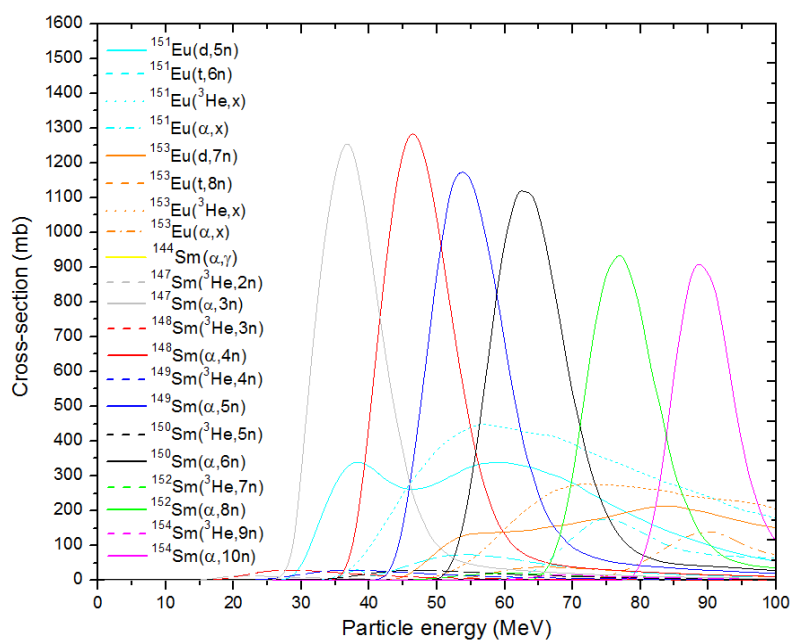

**Fig. 2.** The calculated cross-section curves for the production of Gd-148 on Sm and Eu targets as dependent on projectile energy.

Similarly, other alpha induced reactions have high cross-section values and sharp curves. It has been noted that $^{147}$Sm(α,3n) reaction reaches the maximum point of cross-section curve in low energy (~ 37 MeV alpha incident energy), and the increase emitted neutron number in the reaction processes lead to decreasing cross-section values and increasing incident particle energy. Therefore, $^{154}$Sm(α,10n) reaction has lower cross-section value (~912 mb) and higher incident alpha energy (89 MeV alpha incident energy). Such a circumstance does not demand for the production of radioisotope due to increasing the radioisotope production cost. Nonetheless, the projectile energy in the alpha-induced reaction processes do not exceed 100 MeV and the cross-section values are above 900 mb. On the other hand, the maximum cross-section value of $^{144}$Sm(α,γ) reaction is 0.52 mb at 18 MeV.

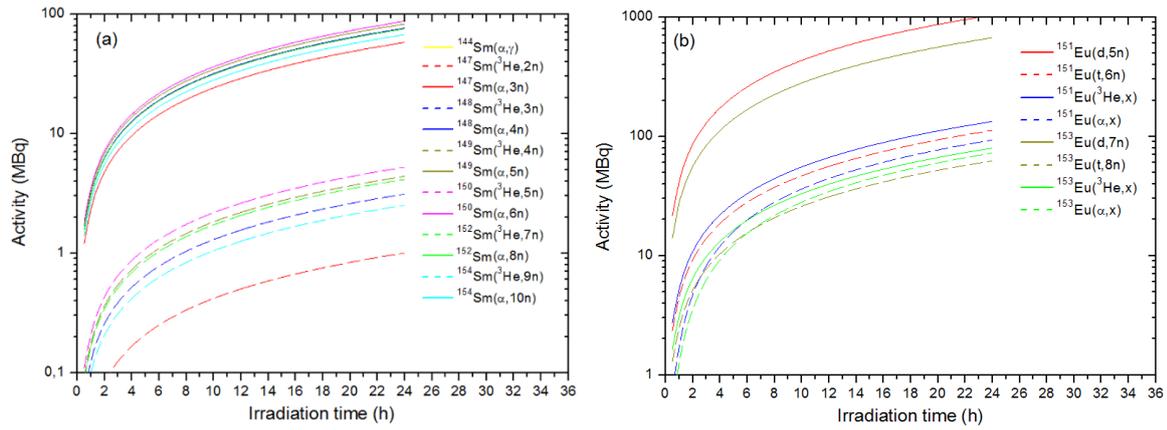

**Fig. 3.** The simulated activitiy curves for the production of Gd-148 on Sm and Eu targets as dependent on irradiation time.

On the other hand, the cross-section curves for the production of Gd-148 on natural Eu isotopes ($^{151,153}$Eu) are wider as compared Sm targets, and contrary to Sm targets, alpha induced reactions on Eu targets have lower cross-section values than those of Sm targets. The cross-sections of $^{151}$Eu($^{3}$He,x) and $^{153}$Eu($^{3}$He,x) reactions for the production of Gd-148 reach the maximum values among Eu targets such as ~449.6 mb at 57 MeV and 275.86 mb at 72 MeV.

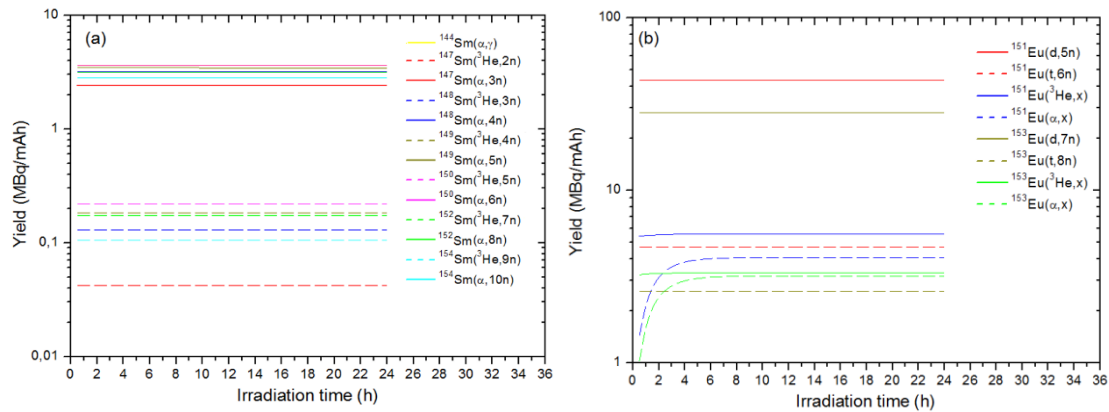

**Fig. 4.** The simulated yield curves for the production of Gd-148 on Sm and Eu targets as dependent on projectile energy.

In analyzing the cross-sections for the production of Gd-148, it is clearly observed that using natural Sm isotopes as target is more appropriate compared to Eu targets. However, to analyze the accuracy of the production of Gd-148, the activities and yields of product of all reaction processes should be also estimated and discussed. In order to estimate the formation of Gd-148 in the charged particle-induced reactions on $^{151,153}$Eu and $^{147,148,149,150,152,154}$Sm target materials, we simulated activity and yield of product of the reaction processes as dependent on irradiation time (Figs. 3 and 4). As can be expected, alpha-induced reactions on Sm targets are consistent with the yield of product and activity curves for the production of Gd-148. However, $^{150}$Sm($\alpha$,6n) and $^{149}$Sm($\alpha$,5n) reach the highest activity values at the end of 24 h irradiation time, 86.80 MBq and 82.44 MBq. Similarly, the yield of product values of these two-reactions also are high as compared to the other reaction processes on Sm targets. On the other hand, the activity and yield of product for $^{148}$Sm($\alpha$,4n) and $^{147}$Sm($\alpha$,3n) reaction processes that correspond to the highest cross-section values are 76.03 MBq, 3.17 MBq/mAh and 56.55 MBq, 2.41 MBq/mAh. As clearly visible, the activity and yield of product values of He-3-induced reactions on Sm targets are lower than those of alpha-induced reactions. In the case of $^{151,153}$Eu targets, based on activity and yield of product, although the cross-section of $^{151}$Eu($^3$He,x) and $^{153}$Eu($^3$He,x) reactions for the production of Gd-148 are higher than those of $^{151}$Eu(d,5n) and $^{153}$Eu(d,7n) reactions, the activities (1036.89 MBq and 670.22 MBq, respectively) for $^{151}$Eu(d,5n) and $^{153}$Eu(d,7n) reactions are higher than the activities of $^{151}$Eu($^3$He,x) and $^{153}$Eu($^3$He,x) reaction processes (133.05 MBq and 79.51 MBq).

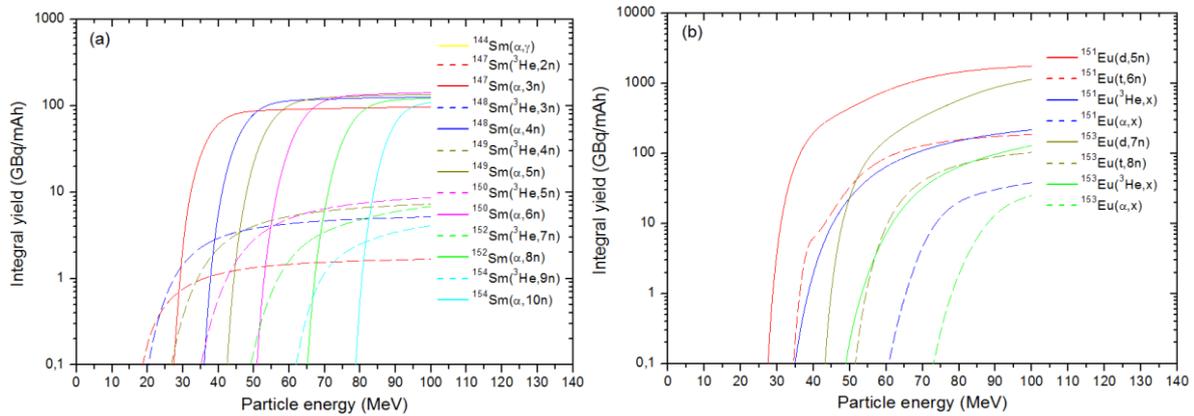

**Fig. 5.** The calculated integral yield curves for the production of Gd-148 on Sm and Eu targets as dependent on projectile energy.

Similar circumstances are available in the yields of product of these two reaction processes, the highest yield of product values are 43.18 MBq/mAh (for $^{151}$Eu(d,5n)) and 27.91 MBq/mAh (for $^{153}$Eu(d,7n)). When taking the activities and the yields of product into account for the formation of Gd-148, contrary to cross-section curves, it has been noted that the activity and yield values on Eu targets are more effective than the reaction processes with the Sm targets up to 10 times. As mentioned above, for determining the appropriate reactions in making up Gd-148, we also calculated the integral yields of the appropriate reaction processes as a function of projectile energy in Fig. 5. In Fig. 5(a), it is obvious that the integral yields of alpha-induced reactions for Sm targets have higher than those of the integral yields He-3-induced reactions. The alpha reactions on Sm targets reach up to 100 GBq/mAh. However, the deuteron-induced reactions on Eu targets in Fig. 5(b) have fairly high integral yield values as compared to the reactions with Sm targets, especially in $^{151}$Eu(d,5n) and $^{153}$Eu(d,7n) reaction processes whose integral yield values are ~1756 GBq/mAh and ~1141 GBq/mAh, respectively. Moreover, the maximum integral yield results of all reaction processes and the suitable energy ranges are presented in Table 1.

## 4. Conclusions

In this work, we analyze the production of Gd-148 because Gd-148 has vital potential as an energy source for use in RTGs. It is obvious that there are ineffective and expensive methods in the literature to produce Gd-148. For this reason, we recommend a new production route of Gd-148 to the literature, which is quite an efficient production method based on the obtained results in this work. The production of Gd-148 is performed by a particle accelerator in the energy region $E_{particle}$=100→1 MeV. The natural Sm and Eu targets are irradiated by the charged particles during 24 hours for a constant beam current of 1 mA. The calculated cross-section and integral yields curves of reaction processes are presented as a function of the projectile energy. Additionally, the activity and the yield curves are simulated for each reaction process as dependent on irradiation time.

When considered the calculated and the simulated results, it has been noted that the importance of target material in reaction processes in the production of Gd-148 can be seen in the obtained results. In terms of the cross-section, the Sm targets are more convenient than those of Eu targets since the cross-section reactions with Sm targets reach above 900 mb at low projectile energies, especially in $^{147}$Sm(α,3n), $^{148}$Sm(α,4n), $^{149}$Sm(α,5n), $^{150}$Sm(α,6n), $^{152}$Sm(α,8n), $^{154}$Sm(α,10n). $^{147}$Sm(α,3n) reaction process has the lowest energy region $E_{particle}$=50→30 MeV and a cross-section value of 1265.61 mb. Unfortunately, for the Sm targets, the $^3$He induced reaction processes do not give the suitable cross-section values for the production of the Gd-148. On the contrary, in the reaction processes with the Eu targets, deuteron, and helium-3 induced reaction processes has the highest cross-section values. However, the projectile energies of these reactions are above 50 MeV. It is obvious that $^{151}$Eu($^3$He,x), $^{153}$Eu($^3$He,x), $^{151}$Eu(d,5n), and $^{153}$Eu(d,7n) reaction processes can be recommended for the production of Gd-148.

On the other hand, if the simulated activity and yield values of the whole reaction processes are taken into account, the processes with the Eu targets is more suitable (about 100 times) than those of the Sm target especially for $^{151}$Eu(d,5n), and $^{153}$Eu(d,7n) processes which the activity and yield values reach 1036.89 MBq, 43.18 MBq/mAh and 670.22 MBq, 27.91 MBq/mAh, respectively.

Based on the whole obtained results, the production of Gd-148 can be performed by a particle accelerator with 100 MeV particle incident energy (in the alpha or deuteron-induced reactions) on the natural Sm and Eu targets, instead of the Gd-148 produced by accelerators with >20 GeV particle incident energy in the literature.